\documentclass[prb,superscriptaddress]{revtex4}
\usepackage{graphicx}
\usepackage{dcolumn}% Align table columns on decimal point
\usepackage{bm}% bold math
\usepackage{amsmath}
\usepackage{comment}
\usepackage{ulem}
\def\ffi{\varphi}
\def\vkk{\frac{1}{|k-k'|^{D-1}}}
\def\Vkk{\frac{1}{|k-k'|}}
\DeclareMathOperator{\asinh}{asinh}
\DeclareMathOperator{\Span}{Span}

\def\simge{\mathrel{ 
       \rlap{\raise 0.511ex \hbox{$>$}}{\lower 0.511ex \hbox{$\sim$}}}}
\def\simle{\mathrel{
       \rlap{\raise 0.511ex \hbox{$<$}}{\lower 0.511ex \hbox{$\sim$}}}}

\begin{document}

\title{Existence of a metallic phase and upper bounds of the Hartree-Fock energy in the homogeneous electron gas}
\author{F. Delyon}
\affiliation{CPHT, UMR 7644 of CNRS, Ecole Polytechnique, Palaiseau, France}
\author{M. Duneau}
\affiliation{CPHT, UMR 7644 of CNRS, Ecole Polytechnique, Palaiseau, France}
\author{B. Bernu}
\affiliation{LPTMC, UMR 7600 of CNRS, Universit\'e P. et M. Curie, Paris, France}
\author{M. Holzmann}
\affiliation{LPTMC, UMR 7600 of CNRS, Universit\'e  P. et M. Curie, Paris, France}
\affiliation{LPMMC, UMR 5493 of CNRS, Universit\'e J. Fourier, Grenoble, France}

\date{\today}
\begin{abstract}
We consider the ground state of the homogeneous electron gas and
we prove that a Hartree-Fock solution, motivated by previous simulations, has lower energy than the Fermi gas in the large density limit.
This solution is a metallic phase : the density modulation corresponds to a partially occupied crystal (the number of sites is larger than the number of electrons).
\end{abstract}
\maketitle
\section{Introduction}
The homogeneous electron gas is one of the fundamental models
in condensed matter physics. Despite its simplicity - the system consists
of electrons interacting with each other through a $1/r$ potential to which a uniform
positive background is added for charge neutrality  - the phase diagram at zero temperature  is 
nontrivial\cite{Tanatar,Bernu,Waintal}. In general, it is given in terms of 
the dimensionless parameter $r_s=(\alpha_D a_B n^{1/D})^{-1}$,
where $D$ is the space dimension, $n$ is the electronic density, $a_B$ the Bohr radius and $\alpha_D^D$ is the volume of the sphere unity. 
% rs = a/a_B  ,  a = 1/( alpha_D n^1/D)
At $r_s=0$, the ground state is the Fermi gas.
At large $r_s$, the Hartree-Fock ground state of the electron gas is a Wigner crystal, that is a state where the charge density  
forms a triangular crystal with
exactly one electron per lattice site. In the intermediate region the Hartree-Fock approximation
is not relevant and more sophisticated methods show that the ground state of the Hamiltonian is quite different.

At small $r_s$, the  ground state of the electron gas within Hartree-Fock
is still not known \cite{Giulani}. 
Although the ground state of the ideal Fermi gas ($r_s=0$) remains an eigenstate
of the Hartree-Fock Hamiltonian of the electron gas at any density, already
Wigner\cite{Wigner}  argued that the unpolarized Fermi gas
is unstable even in the limit $r_s \to 0$.
Later, Overhauser showed the instability of the 
unpolarized Fermi gas with respect to spin-density waves within the Hartree-Fock
approximation\cite{Overhauser}.
Only recently, a Hartree-Fock study 
of the unpolarized three-dimensional electron gas was performed which proposes a more
complicated structure of a ground state with spin-density waves\cite{Shiwei}. 

In Ref. \cite{BDDH}, we describe the results of our numerical simulations of the two-dimensional (2D), polarized electron gas at small $r_s$, and show that the ground state is neither a Fermi gas nor a Wigner crystal: the charge density modulation we find corresponds to a partially occupied crystal (the number of sites is larger than the number $N$ of electrons).
We refer to this solution as a metallic phase. The charge density of this metallic phase represents a triangular lattice with
reciprocal generators $Q_i$ of modulus $2k_F$. This modulation is mainly carried by the wave vectors close to the Fermi surface. In order to observe the metallic phase, the number of electrons $N$ has to exceed a threshold 
ranging from 
$N \simge 10$  at $r_s\sim 2.6$ up to $N \simge 10^2$ at $r_s\sim 1$.
As we shall see below, 
the threshold increases exponentially as $r_s$ goes to 0
which may explain why this metallic phase has not been observed in previous simulations.

In this study, we obtain rigorous upper bounds on the energy of the metallic phase in the two-dimensional polarized case with a 
class of states mimicking the real states obtained numerically. These bounds are obtained in the limit $r_s \to 0$ where the calculation is simplified by the long range behavior
of the interaction potential.
However, 
 our numerical simulations show that this metallic phase also exists for screened potential, but strictly speaking 
we cannot prove in this case the existence of such a metallic phase in the limit $r_s\to0$.
This bounds are easily extended to the unpolarized 2D case.
Then we show how this proof extends to the 3D case, whereas we have no numerical simulations indicating that this phase may 
correspond to  the actual ground state of the 3D electron gas in Hartree-Fock.

Finally, let us point out that our bounds (for instance Eq.~(\ref{deb})) must be considered as mathematical bounds; they only give 
an understanding of the behavior of the energy of the ground state,
but the actual constants involved in the expressions have to be evaluated by other means.

\section{Numerical results}
In a previous paper\cite{BDDH},
we have computed 2D Hartree-Fock states of lower energy than the Fermi gas for values of $r_s$ about 2.

\begin{figure}
\begin{center}
\includegraphics[width=0.4\textwidth]{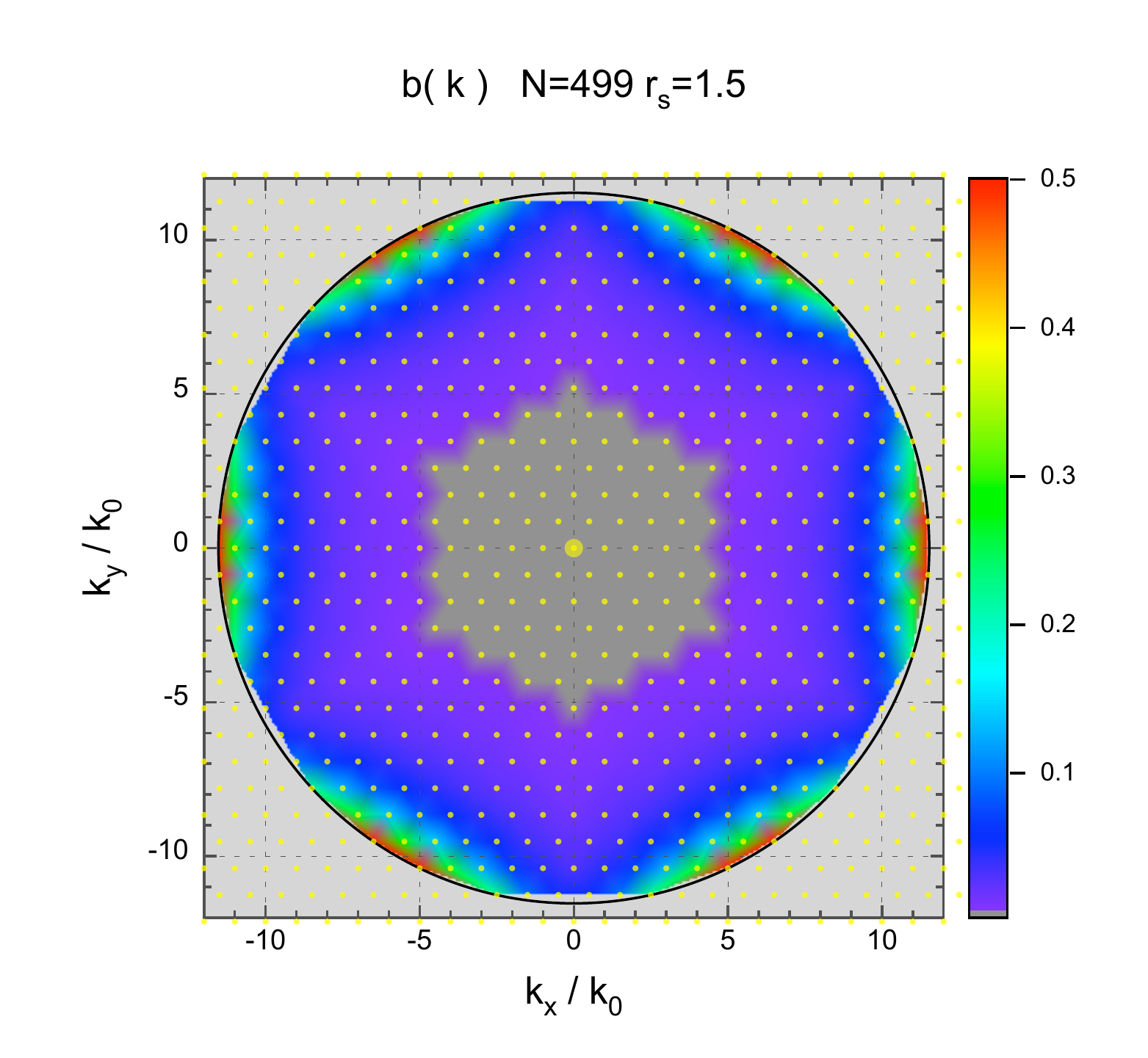}
\caption{Numerical values of $b_k$ for 499 electrons in two dimensions at $r_s=2$.
}
\label{b499}
\end{center}
\end{figure}

Such a state $\Psi$ is obtained as a Slater determinant $\bigwedge_{i=1}^N \psi_i$ where $N$ is the number of electrons and $\{\psi_i\}$ is a set of orthonormal vectors.
Only the space generated by the single particle wavefunctions $\psi_i$'s is relevant, and in order to understand our numerical results we need to choose a canonical representation of the $\psi_i$'s.\\
The natural choice is to use the natural base of the Fermi gas $\{\phi_i\}_{i=1\ldots N}$ corresponding to some indexation of the plane waves of the Fermi sphere.\\
Let $M$ be the square matrix defined by $M_{ij}=\left<\phi_i| \psi_j\right>$. The Singular Value Decomposition (SVD)  of $M$ is $M=U\sigma V$ where $U$ and $V$ are unitary matrices and $\sigma$ is a diagonal positive matrix.\\
Let us define 
\begin{eqnarray}
\phi_i'&=&\sum_j \overline U_{ij}^{-1}\phi_j\\
\psi_i'&=&\sum_jV_{ji}^{-1}\psi_j .
\end{eqnarray}
Then  $\{\phi_i'\}$ and $\{\psi'_i\}$ are also two sets of $N$ orthonormal vectors  satisfying $\left<\phi_i'|\psi_j'\right>=\delta_{ij}\sigma_i$.\\
They are the nearest bases of $\Span(\{\phi_i\})$ and  $\Span(\{\psi_i\})$ and they define a canonical unitary operator $W$ from $\Span(\{\phi_i\})$ onto  $\Span(\{\psi_i\})$ by:
\begin{eqnarray}
W\sum_i \lambda_i\phi_i'&=\sum_i \lambda_i\psi_i'
\end{eqnarray}
In particular, we have:
\begin{eqnarray}
W\phi_i&=&W \sum_j \overline U_{ij}\phi_j'\\
&=&\sum_j \overline U_{ij}\psi_j'\\
&=&\sum_j \overline U_{ij}\sum_kV_{kj}^{-1}\psi_k\\
&=&\sum_j \overline{UV}_{ik}\psi_k
\end{eqnarray}
Thus the natural basis of $\Span(\{\psi_i\})$ associated to the basis $\{\phi_i\}_{i=1\ldots N}$ is $\{W\phi_i\}_{i=1\ldots N}$.\\
Numerically we have chosen the $\psi_i$'s in this way and it appears that, if $r_s$ is not too large, $\psi_i$ is close to $\phi_i$ at least for $i$ associated to a 
wave vector not too close to 
the Fermi surface. Thus the largest amplitude of $\psi_i$, in the $k$ space, is for the $k_i$ corresponding to $\phi_i$.
Fig.\ref{b499} represents the next largest amplitude of $\psi_i$, that we denote $b_{k_i}$, for 499 electrons at $r_s=2$ in 2D.

As $r_s$ decreases, the other amplitudes of the $\psi_i$'s become very small.
Furthermore, the wave vector $k'_i$ corresponding to $b_{k_i}$ satisfies $k'_i-k_i=Q_\alpha$ where the generators $\{Q_\alpha\}_{\alpha=1\ldots 6}$ belong to the six-fold star of a triangular lattice. 

Notice that the modulus of $b_k$ is maximal for $k$ on a  six-fold star and close to the Fermi surface.
In the direct space,  the $Q_{\alpha}$'s induce a modulation corresponding to a triangular lattice for the charge density.
But while this lattice forms a Wigner crystal with one electron per site for $r_s>2.6$,
for smaller $r_s$ it is a denser lattice corresponding to a crystal with an incomplete band filling.
Indeed, at any $r_s$ the electrostatic interaction favors a periodic distribution of 
charges and at small $r_s$ only $k$-vectors close to Fermi surface can be modulated.
The optimal solution is to choose $\|Q_\alpha\|=2k_F$ which is larger than the
reciprocal lattice vector of the Wigner case.
In the following, we consider analytic solutions analogous to those of our numerical results and we prove that  their energies are lower than the energies of the Fermi
gas as $r_s$ goes to 0.

\section{Energy of the Fermi gas: polarized case}
We consider the Hamiltonian of $N$ electrons in a 2D or 3D square box of volume $\Omega$ with periodic boundary conditions.\\
\begin{align}
H&=-\frac{\hbar^2}{2m}\Delta+\frac{e^2} 2  V
\end{align}
where $V$ is the 2-body Coulomb potential $\sum_{i\neq j}1/|r_i-r_j|$, 
 the electron mass is $m$, and $e$ is its charge.
It is convenient to choose Hartree as the unit of energy,
$Ha=\hbar^2/(ma_B^2)$, where $a_B=\hbar^2/(me^2)$ is the Bohr radius.
We get:
\begin{align}
H&=\frac{a_B^2}2 (-\Delta+ \frac 1 {a_B}V)
\end{align}
Let $\psi_n$ be an orthonormalized set of $N$ vectors of $L^2(\Omega)$. They define the $N$-particle
Slater determinant $\Psi=\bigwedge_{n} \psi_n$. And the energy of $\Psi$ is:
\begin{align}
\label{energy}
{\cal E}=\left<\Psi | H | \Psi\right>&=\frac{a_B^2}{2} \left(-\sum_{n}\left<\psi_n| \Delta | \psi_n\right>+ \frac {1} {a_B}\sum_{n,n'}\left<\psi_n\wedge\psi_{n'} | v | \psi_n\wedge\psi_{n'}\right>\right)
\end{align}
where $v$ is defined as:
\begin{align}
\left<\ffi_1\otimes\ffi_2|v|\psi_1\otimes\psi_2\right>&=
\int dx\,dy \ \overline{\ffi_1}(x) \overline{\ffi_2}(y)\frac{1}{||x-y||}\psi_1(x)\psi_2(y).
\end{align}
In order to avoid problems due to the Coulomb singularity, we
 introduce the jellium model and  define the potential acting
on the plane waves $\phi_k$ as:
\begin{align}
\left<\phi_k\otimes\phi_{k'}|v|\phi_{k+q}\otimes\phi_{k'-q}\right>&=\frac{\pi}{\Omega}\left(\frac{2}{|q|}\right)^{D-1}
\end{align}
for $q\neq 0$ and $0$ otherwise, so that the total charge of the electrons is compensated
by a positive background charge.\\
The Fermi gas is  defined by $\Phi=\bigwedge_{|k|<k_F} \phi_k$ where $(\alpha_Dk_F^D)=(2\pi)^D N/\Omega$ and $\alpha_D^D$ is the volume of the unit sphere.
\begin{align}
{\cal E}_{FG}=\left<\Phi|H|\Phi\right>&=\frac{a_B^2}{2} \left(\sum_{|k|<k_F}k^2- \frac {2^{D-1}\pi} {a_B\Omega}\sum_{|k|,|k'|<k_F}\vkk\right)
\end{align}
As $\Omega$ goes to $\infty$ with  $\Omega/N$ fixed, the thermodynamic
limit for the energy per particle is obtained by the substitution $\sum_k \to \frac{\Omega}{(2\pi)^D} \int dk$:
\begin{align}
\nonumber
\frac{{\cal E}_{FG}}N&=\frac{a_B^2}2 \frac{\Omega}{N(2\pi)^D}\left(\int_{|k|<k_F}dk\ k^2- \frac 1 {a_B2\pi^{D-1}}\int_{|k|,|k'|<k_F}dkdk'\ \vkk\right)\\
&=\frac{a_B^2}2 \frac{\Omega}{N(2\pi)^D}k_F^{D+2}\left(\int_{|k|<1}dk\ k^2- \frac 1 {a_Bk_F2\pi^{D-1}}\int_{|k|,|k'|<1}dkdk'\ \vkk\right)
\end{align}
From the definition of $r_s=(\alpha_D a_B n^{1/D})^{-1}$ and $k_F$, it follows that $k_F\alpha_D^2r_s a_B=2\pi$.
%, thus $r_sa_Bk_F=2$.
Thus, we have:
\begin{align}
\frac{{\cal E}_{FG}}N=\frac{2\pi^2}{\alpha_D^{D+4} r_s^2}\left(\int_{|k|<1}dk\ k^2- \frac {r_s\alpha_D^2} {4\pi^D}\int_{|k|,|k'|<1}dkdk'\ \vkk\right)
\end{align}
which gives for $D=2$  ($\alpha_2^2=\pi$):
\begin{align}
\label{efg}
\frac{{\cal E}_{FG}}N=\frac{2}{\pi r_s^2}\left(\int_{|k|<1}dk\ k^2- \frac {r_s} {4\pi}\int_{|k|,|k'|<1}dkdk'\ \Vkk\right)
\end{align}
\section{Hartree-Fock upper bounds: polarized 2D case}
\begin{figure}
\begin{center}
\includegraphics[width=0.4\textwidth]{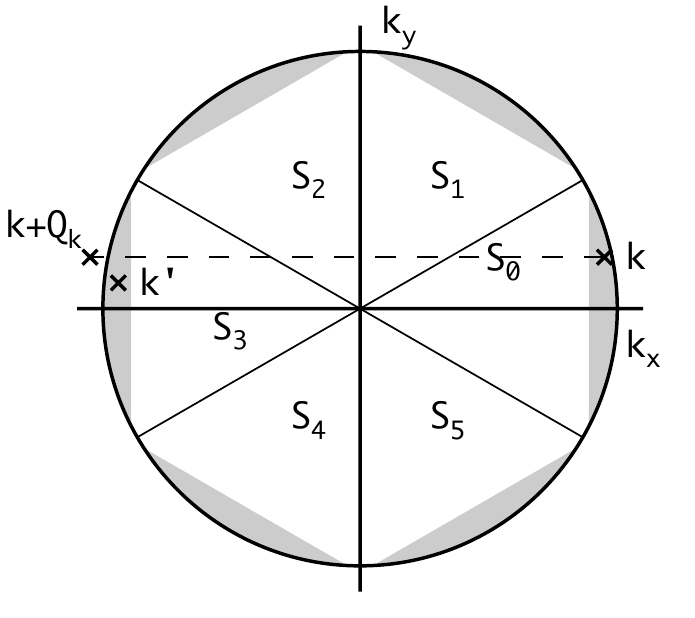}
\caption{The circle is the Fermi surface. The shaded surfaces are the regions where $b(k)$ is nonzero. The new state $\psi_k$
mixing $\phi_k$ and $\phi_{k+Q_k}$ is now resonant with $\psi_{k'}$.
}
\label{resonant}
\end{center}
\end{figure}

We restrict ourself to the 2D polarized case and we want to estimate the energy for a class of states inspired by our numerical results.
Let us consider a state $\Psi=\bigwedge_{|k|<k_F} \psi_k$
where:  $$\psi_k=a_k\phi_k+b_k\phi_{k+Q_k}$$
with $Q_k$ in $\{-2k_F(\cos p\pi/3,\sin p\pi/3)\}_{p=0\ldots 5}$.
For $k=|k|(\cos \theta, \sin \theta)$ we choose $Q_k$ such that $|k+Q_k|$ is minimal; that is, we choose $p$ as the integer part of $(3\theta/\pi+1/2)$ and
 we must assume $b_k$ is zero if $k$ is zero or $\theta=\pi/6+n\pi/3$.\\
Furthermore, we assume that $a_k$ and $b_k$ are real positive number and invariant thru the rotation of $2n\pi/6$ and the symmetry $\theta\rightarrow -\theta $ 
(i.e. the dihedral group $D_6$). The $ \psi_k$'s are normalized, so that $a_k^2+b_k^2=1$ and  $b_k=0$ if $|k\cdot Q_k|<2k_F^2(1-\epsilon)$ 
(i.e. $b_k$ is not zero only in the vicinity of $\{k_F(\cos p\pi/3,\sin p\pi/3)\}_{p=0\ldots 5}$), see Fig.\ref{resonant}. \\
Thus the limit energy per particle is given by: 
\begin{align}
\frac{{\cal E}}N=\frac{2}{\pi r_s^2}\left(\int_{|k|<1}dk\ \left<\psi_k|-\Delta | \psi_k\right>+ \frac {r_s} {4\pi}\int_{|k|,|k'|<1}dkdk'\ \frac{\Omega}{2\pi}\left<\psi_k\wedge\psi_{k'}|v| \psi_k\wedge\psi_{k'}\right>\right)
\end{align}
where, as in (\ref{efg}),  the $k$'s have been renormalized by $k_F$ and thus $|Q_k|=2$.

We define $\Delta E$ by:
\begin{align}
\label{deltaEdef0}
\frac{{\cal E}-{\cal E}_{FG}}N=\frac{2}{\pi r_s^2}\Delta E
\end{align}
Then
\begin{align}
\label{defde}
\Delta E=\int_{|k|<1}dk\,\left[\left<\psi_k|-\Delta| \psi_k\right>-k^2\right]+ \frac {r_s} {4\pi}\Delta E_V
\end{align}
where
\begin{align}
\label{dev}
\Delta E_V=\int_{|k|,|k'|<1}dkdk'\, \left(\frac{\Omega}{2\pi}\left<\psi_k\wedge\psi_{k'}|v | \psi_k\wedge\psi_{k'}\right>+\Vkk\right)
\end{align}\\
\subsection{Potential energy contribution: $\Delta E_V$}
Setting $v_q=1/|q|$:
\begin{align}
\nonumber
\frac{\Omega}{2\pi}\left<\psi_k\wedge\psi_{k'}|v| \psi_k\wedge\psi_{k'}\right>+v_{k-k'}&=(v_{k-k'}-v_{k-k'-Q_{k'}})b_{k'}^2a_k^2\\
\nonumber
&+(v_{k-k'}-v_{k+Q_{k}-k'})b_{k}^2a_{k'}^2\\
\nonumber
&+(v_{k-k'}-v_{k+Q_{k}-Q_{k'}-k'})b_{k}^2b_{k'}^2\\
\nonumber
&+2v_{Q_k}a_k b_k  a_{k'} b_{k'}(\delta_{Q_k+Q_{k'}}+\delta_{Q_k-Q_{k'}})\\
\nonumber
&-2v_{k-k'}a_k b_k  a_{k'} b_{k'}\delta_{Q_k-Q_{k'}}\\
\label{ve0plusv}
&-(v_{k+Q_k-k'}+v_{k-k'-Q_{k'}})a_{k} a_{k'}  b_k b_{k'}\delta_{Q_k+Q_{k'}}
\end{align}
Eq.(\ref{dev}), may be divided into 4 parts:
\begin{itemize}
\item \underline{\{$b_{k'}=0$, $b_k= 0$\}:}
the contribution is zero. 
\item \underline{\{$b_{k'}=0$, $b_k\neq 0$\}, \{$b_{k'}\neq 0$, $b_k= 0$\}:}
both cases are equivalent. 

For \{$b_{k'}=0$, $b_k\neq 0$\}, the integrant of Eq. \ref{dev} is:
%Let us consider the case \{$b_{k'}=0$, $b_k\neq 0$\} (the case \{$b_{k}=0$, $b_{k'}\neq 0$\} is equivalent):
\begin{align}
\frac{\Omega}{2\pi}(\psi_k\wedge\phi_{k'},v\ \psi_k\wedge\phi_{k'})+\Vkk=(v_{k-k'}-v_{k+Q_{k}-k'})b_k^2
\end{align}
Let  $S_0$ be the sector of unit disk between $-\pi/6$ and $\pi/6$ (see Fig. \ref{resonant}); then in this sector $Q_{k}=(-2,0)$ and by symmetry:
\begin{align}
\int_{b_{k'}=0}dkdk'\ (v_{k-k'}-v_{k+Q_{k}-k'})b_k^2
&=6\int_{k\in S_0, b_{k'}=0}dkdk'\ (v_{k-k'}-v_{k+Q_{k}-k'})b_k^2\\
&=6\int_{k\in S_0, b_{k'}=0}dkdk'\ (v_{k-k'}-v_{\tilde k-k'})b_k^2\\
&\leq C\epsilon^3+6\int_{k\in S_0, |k'_x|<1-\epsilon}dkdk'\ (v_{k-k'}-v_{\tilde k-k'})b_k^2
\end{align}
where $\tilde k=(2-k_x,k_y)$.
In $S_0$, $k=(k_x,k_y)$ where $k_x$ is close to 1 and setting $k_x=1-x$, we 
assume from now that $b_k=b(x/\epsilon)$.\\
In Appendix A we prove that:
\begin{align}
\label{bk0}
\int_{k\in S_0,|k'_x|<1-\epsilon}dkdk'\ (v_{k-k'}-v_{k+Q_{k}-k'})b_k^2\leq 8\epsilon^2\sqrt{2\epsilon}\left[\ln\epsilon^{-1}+O(1)\right]\int_0^1 dx b^2(x)x\sqrt x
\end{align}
\item \underline{\{$b_{k'}\neq0$, $b_k\neq 0$\}:}

By symmetry we can assume that $k$ belongs to $S_0$. If $k'\not \in S_0\cup S_3$ all the $v$ appearing in (\ref{ve0plusv}) are uniformly  bounded.
And since the k-volume for each sector goes like $\epsilon\sqrt\epsilon$, the contribution of these terms is bounded by $C\epsilon^3$.
 In the same way $v_{k-k'}$ is bounded when $k'\in S_3$ and $v_{k+Q_k-k'}$ is bounded when $k'\in S_0$.\\
 Thus setting:
\begin{align}
f&:=a_k^2b_{k'}^2+b_k^2a_{k'}^2-2a_{k} a_{k'}  b_k b_{k'}=(a_kb_{k'}-b_k a_{k'})^2\\
g&:=a_k^2b_{k'}^2+b_k^2a_{k'}^2+2a_ka_{k'}b_{k} b_{k'}=(a_kb_{k'}+b_k a_{k'})^2
\end{align}
one can check that:
\begin{align}
\label{dev2}
\int_{b_k,b_{k'}\neq 0}\frac{\Omega}{2\pi}\left<\psi_k\wedge\psi_{k'}|v| \psi_k\wedge\psi_{k'}\right>+v_{k-k'}\leq C\epsilon^3+6\int_{k,k'\in S_0}dkdk'\ \left(v_{k-k'}f-v_{k+Q_k+k'}g\right)
\end{align}
In Appendix B we prove that
\begin{align}
\label{bk1}
\int_{k,k'\in S_0}dkdk'\ \left(v_{k-k'}f-v_{k+Q_k+k'}g\right)\leq 4\epsilon^2\sqrt{2\epsilon}\left[\ln\epsilon^{-1}+O(1)\right]\int_0^1 dx\sqrt x\int_x^1 dx' \left(f( \epsilon x, \epsilon x')-g( \epsilon x, \epsilon x')\right)
\end{align}
\end{itemize}
Thus, summing the four contribution gives:
\begin{align}
\label{dev3}
\Delta E_V&\leq C\epsilon^3+6\epsilon^2\sqrt{2\epsilon}\left[\ln\epsilon^{-1}+O(1)\right]\int_0^1 dx\sqrt x\left(16b^2(x)x+4\int_x^1 dx'\ \left(f(x, x')-g( x, x')\right)\right)
\end{align}

\subsection{Kinetic energy contribution:}
The variation of the kinetic energy is given by:
\begin{align}
\int_{|k|<1}dk\ \left[\left<\psi_k | -\Delta | \psi_k\right>-k^2\right]&=6\int_{k\in S_0} dk (\left<\psi_k|-\Delta| \psi_k\right>-k^2)\\
&=6\int_0^\epsilon dx\ 2y_m\ 4xb^2(x/\epsilon)\\
\label{kin}
&\leq 6\times 8\epsilon^2\sqrt{2\epsilon}\int_0^1 dx\ \sqrt x\ xb^2(x)
\end{align}

\subsection{Total energy:}
Inserting Eqs.(\ref{dev3},\ref{kin}) in Eq. (\ref{defde}), the variation of the total energy from the Fermi gas energy becomes:
\begin{align}
\nonumber
\Delta E&\leq 6\epsilon^2\sqrt{2\epsilon}\int_0^1 dx\sqrt{x}\left(8xb^2( x)+\frac {r_s} {4\pi}\left[\ln\epsilon^{-1}+O(1)\right]\left(16b^2( x)x+4\int_x^1 dx'\left(f(x, x')-g(x, x')\right)\right)\right)\\
\label{def}
&= 6\times 8\epsilon^2\sqrt{2\epsilon}\int_0^1 dx\sqrt{x}\left(xb^2( x)+\frac {r_s} {2\pi}\left[\ln\epsilon^{-1}+O(1)\right]\left(b^2( x)x-a(x)b(x)\int_x^1 dx'a(x')b(x'))\right)\right)
\end{align}
Let us set 
\begin{align}
\label{defdelta}
\delta&=\epsilon^2\sqrt{\epsilon}\\
\label{i1i2-1}
I_1&=\int_0^1 dx\sqrt xxb^2( x)\\
\label{i1i2bis-1}
I_2&=\frac{1}{5\pi}\int_0^1 dx\sqrt x\left(-b^2( x)x+a(x)b(x)\int_x^1 dx'a(x')b(x')\right)
\end{align}
Then
\begin{align}
\label{deltaEdef}
\Delta E&\leq  6\times 8\sqrt{2}\delta\left[I_1-r_sI_2(\ln \delta ^{-1}+O(1))\right]
\end{align}
If $I_2>0$, as $r_s$ goes to $0$, $\Delta E$ is minimal in Eq. (\ref{deltaEdef}) for  $\delta$ defined by:
\begin{align}
\label{deltamin}
\delta_{\min}=\frac 1 {e}\exp\left(-\frac {I_1}{I_2r_s}\right)
\end{align}
and finally inserting $\delta_{\min}$ in Eq. (\ref{deltaEdef}) gives:
\begin{align}
\label{DEF1}
\Delta E\lesssim-\frac{6\times 8\sqrt{2}}{e}\exp\left(-\frac {I_1}{I_2r_s}\right)r_sI_2
\end{align}
We now have to find a solution $b(x)$ such that $I_2$ is positive. Choosing $b(x)=b_0$ or $b(x)=b_0 ( 1 - x ) $ leads to negative $I2$. 
In the Appendix C, as $r_s$ goes to 0 we find a family of $b$ leading to :
\begin{align}
\label{deb}
\Delta E&\lesssim -r_s\exp\left(-\frac {5\pi}{3r_s}+\frac {O(1)} {\sqrt {r_s}}\right)
\end{align}
Though such a bound is correct in the thermodynamic limit, this behavior in not so relevant for finite systems.
Indeed, numerical systems consider about $10^3$ electrons and, as we shall see later, the Fermi gas becomes then the ground state for $r_s \lesssim 1$.
Thus the asymptotic bound (\ref{deb}) is not very helpful in the real word.\\
Nevertheless, for realistic $r_s$, on can choose a suitable function $b$ and evaluate numerically $I_1$ and $I_2$.
For instance, with $b=b_\eta$ as in (\ref{beta}) of Appendix C and $\eta=0.001$ we get 
\begin{align}
\Delta E&\lesssim-2.6\times 10^{-4}r_s\exp\left(-\frac { 18.5}{r_s}\right)
\end{align}

\section{HF Upper bounds: unpolarized 2D case}
We consider now the spin of the electrons, and we restrict to the case where half the electrons have a spin up and the others have a spin down.\\
We can choose a solution as the product of 2 Slaters $\Psi^+$ and $\Psi^-$ for the spins up and down.
\begin{align}
\label{energy2}
\frac{2}{a_B^2}{\cal E}^{NP}=&-\sum_{n}\left<\psi_n^+|\Delta|\psi_n^+\right>+ \frac {1} {a_B}\sum_{n,n'}\left<\psi_n^+\wedge\psi_{n'}^+|v|\psi_n^+\wedge\psi_{n'}^+\right>\\
\label{energy21}
&-\sum_{n}\left<\psi_n^-|\Delta|\psi_n^-\right>
+ \frac {1} {a_B}\sum_{n,n'}\left<\psi_n^-\wedge\psi_{n'}^-|v|\psi_n^-\wedge\psi_{n'}^-\right>\\
\label{energy22}
&+ \frac {1} {a_B}\left(\sum_{n,n'}\left<\psi_n^+\otimes\psi_{n'}^-|v|\psi_n^+ \otimes\psi_{n'}^-\right>+\sum_{n,n'}\left<\psi_n^-\otimes\psi_{n'}^+|v|\psi_n^-\otimes\psi_{n'}^+\right>\right)
\end{align}
The first terms (\ref{energy2}) and  (\ref{energy21}) can be evaluated as in (\ref{deltaEdef0}) noticing that the definition of $k_F$ becomes $k_F^2=4\pi N_\pm/\Omega=2\pi N/\Omega$. So 
the $r_s$ in the evaluation of $\Delta E$ is now $r_s\sqrt 2$ and:
\begin{align}
\frac{{\cal E}^{NP}-{\cal E}_{FG}^{NP}}{N}=\frac{1}{\pi r_s^2}\Delta E(r_s\sqrt 2)+(\ref{energy22})
\end{align}
The extra terms (\ref{energy22}) coming from the direct potential provide contributions like $a^+_nb^+_na^-_{n'}b^-_{n'}$ times a positive factor. These contributions are 
regular as $r_s$ goes to $0$, and does not modify the asymptotic energy.\\
Nevertheless, these contributions depends on
the signs of $b^\pm_n$ in the two Slaters $\Psi^+$ and $\Psi^-$.
Thus, since the energy of one Slater does not depend on the global sign of the $b^\pm_n$'s, the minimal state is obtained by choosing $b^-_n=-b^+_n$.

For such a state, the main difference with the polarized case is that the charge density becomes flat, though the spin density is not.

Finally, we can compare our solutions with previous solutions proposed by Fedders and Martin\cite{Fedders}.
They consider a more complicated modulation of the states near the center of Fermi sphere. While their results may be correct,
they missed the dominant contribution coming from the states near the Fermi surface. Indeed, our computations give nonzero contribution for the
states near the center, but for $r_s=1.5$ $b_k$ is about $10^{-5}$ and the energy benefit,
following their formula, must be of order $10^{-20}$ of the energy  benefit of the external states.

\section{The metallic phase in finite systems}
For finite systems of $N$ electrons, the minimum requirement is that the surface 
$|k \cdot Q_k|>2k_F^2(1-\epsilon)$, so that it contains at least one plane wave of the
 finite, discrete system.
This gives the condition
$N\epsilon\sqrt\epsilon>1$ and from Eqs. (\ref{defdelta},\ref{deltamin}), this leads to:
\begin{align}
N>\exp\left(\frac{3I_1}{5I_2 r_s}\right)
\end{align}
With our approximate behavior for $I_2/I_1$ in two dimensions, we find:
\begin{align}
N>\exp\left(\frac{3 \pi}{ r_s}\right)
\end{align}
i.e. $N>500$ for $r_s=1.8$. This bound is compatible with our numerical simulations where the metallic phase disappears at $rs=1$ for $N=500$.
Furthermore, this may explain why this metallic phase has not been observed in previous simulations.

\section{Hartree-Fock upper bounds: polarized 3D case}
Mutatis mutandis, the 3D case works in the same way.\\
The potential term $v_k$ is now $1/\|k\|^2$ but the dominant terms are provide now from integrals like:
\begin{align}
I(k_x,k_{x'})=\int_{\|k\|,\|k'\|\leq 1}dk_ydk_zdk'_ydk'_z\ \frac 1 {\|k-k'\|^2}
\end{align}
where $k_x$ is close to $k'_x$.\\
Let us suppose that $k_x>k'_x$, then 
\begin{align}
I(k_x,k_{x'})=\int_0^{\sqrt{2-k_x^2-{k'}_x^2}} \mu(du)\ \frac 1 {(k_x-k'_x)^2+u^2}
\end{align}
where $u$ stands for $((k_y-k'_y)^2+(k_z-k'_z)^2)^{1/2}$ and $\mu(du)$ stands for the distribution of $u$.
Here we are interested in small values of $u$.
One can check that $\mu(u,u+\epsilon)\approx2\pi^2 (1-k_x^2)\epsilon u$ provided that $u$ is sufficiently small ($(k_y, k_z)$ may 
be any point in disk of radius $\sqrt{1-k_x^2}$ and $(k'_y, k'_z)$ may belong to a thin ring around $(k_y, k_z)$ )
\begin{align}
I(k_x,k_{x'}&)\approx \pi^2 (1-k_x^2)\int_0^{2\sqrt{2\epsilon}} 2udu\ \frac 1 {(k_x-k'_x)^2+u^2}\\
&\approx -\pi^2 (1-k_x^2)\ln (k_x-k'_x)^2
\end{align}
This must be compared with the 2D case $
%\begin{align}
I(k_x,k_{x'})%&=\int_{\|k\|,\|k'\|\leq 1}dk_ydk'_y\ \frac 1 {\|k-k'\|}\\
%&
\approx -2\sqrt{1-k_x^2}\ln (k_x-k'_x)^2
%\end{align}
$.
Mutatis mutandis, (Eq. \ref{defde}) becomes:
\begin{align}
\label{defde3}
\Delta E=\int_{|k|<1}dk\ \left(\left<\psi_k | -\Delta | \psi_k\right>-k^2\right)+ \kappa r_s\Delta E_V
\end{align}
where $\kappa= \alpha_3^2/(4\pi^3)^2$ and $\alpha_3^3 = 4\pi/3$. Now, in $\Delta E_V$ the $\sqrt{2\epsilon x}$ factors have to be replaced by $\pi^2\epsilon x$:
\begin{align}
\label{dev33}
\Delta E_V&\leq C\epsilon^4+16z\pi^2\epsilon^3\left[\ln\epsilon^{-1}+O(1)\right]\int_0^1 dx\ x\left(b^2(x)x- a(x)b(x)\int_x^1 dx'a(x')b(x')\right)
\end{align}
where $z$ is the number of sectors of the Fermi sphere.
The variation of the kinetic energy is now
\begin{align}
\int_{|k|<1}dk\ \left(\left<\psi_k|-\Delta|\psi_k\right>-k^2\right)=z8\pi\epsilon^3\int_0^1 dx\ x^2b^2(x)
\end{align}
Thus setting
\begin{align}
\label{i1i2-2}
I_1&=\int_0^1 dx\ x^2xb^2( x)\\
\label{i1i2bis-2}
I_2&=\frac{2\kappa \pi} 3\int_0^1 dx\ x\left(-b^2( x)x+a(x)b(x)\int_x^1 dx'a(x')b(x')\right)
\end{align}
\begin{align}
\nonumber
\Delta E&\leq  z8\pi\epsilon^3\left(I_1-3r_sI_2\left[\ln \epsilon ^{-1}+O(1)\right]\right)
\end{align}
Once more, if we find a $b$ such that $I_2$ is positive, we have:
\begin{align}
\label{DEF2}
\Delta E\lesssim-\frac{z8\pi}{e}\exp\left(-\frac {I_1}{I_2r_s}\right)r_sI_2
\end{align}
The operator $A$ of Appendix C is now modified. But the same analysis show that the spectrum of $A$ is still $(0,4)$, so that the asymptotic bound for small $r_s$ is:
\begin{align}
\Delta E\lesssim-\frac{z8\pi}{e}\exp\left(-\frac {1}{6\kappa\pi r_s}+o(1/r_s)\right)r_s
\end{align}
\section{Conclusion}
We have proven upper bounds for the energy of a metallic state below the Fermi gas energy. 
To our knowledge, it is the first time that rigorous  upper bounds 
for the ground state energy of the polarized electron gas are obtained going beyond the
simple Fermi gas states.
The main interest of this proof is
that it focuses on the small $r_s$ region where the Hartree-Fock approximation 
is relevant 
and that
the upper bounds were obtained using metallic states which  correspond to 
solutions obtained by numerical simulations. 

In order to judge their relevance
for the true ground state of the electron gas in the high density region, we have to
consider correlation effects beyond the Hartree-Fock approximation. Due to the 
long-range, singular behavior of the Coulomb-potential, the perturbation expansion
has to be rearranged which amounts to an effective screening of the electron interaction.
Unfortunately  a rigorous extension of our proof to the case of
a screened Coulomb potential is not straightforward. However, numerical calculations provide evidence
for the stability of the metallic states in the 2D polarized system within Hartree-Fock, 
so that these states are relevant candidates for the true ground state of the
two-dimensional electron gas.

\section{Appendix A}
We have to estimate 
\begin{align}
I(f)=\int_{\begin{subarray}{1} \|k\|,\|k'\|\leq 1\\1-k_x<\epsilon,|k'_x|<1-\epsilon\end{subarray}}dkdk'\ (v_{k-k'}-v_{\tilde k-k'})f(1-k_x)
\end{align}
where $\tilde k=(2-k_x,k_y)$ and $f$ is
a positive function.
\begin{align}
\int dk'\ (v_{k-k'}-v_{\tilde k-k'})=\int dk_x'\ \asinh\frac{y'_m-k_y}{k_x-k'_x}+\asinh\frac{y'_m+k_y}{k_x-k'_x}- \asinh\frac{y'_m-k_y}{2-k_x-k'_x}-\asinh\frac{y'_m+k_y}{2-k_x-k'_x}
\end{align}
where $y'_m=\sqrt{1-{k'_x}^2}$. And since $\asinh x -\asinh y\leq\ln x/y$ for $x>y>0$:
\begin{align}
\int dk'\ (v_{k-k'}-v_{\tilde k-k'})\leq\int_{-1+\epsilon}^{1-\epsilon} dk_x'\  2\ln\frac{2-k_x-k'_x}{k_x-k'_x}
\end{align}
We set $k_x=1-x$ and $y_m=\sqrt{2x-x^2}$,
\begin{align}
I(f)&\leq\int_0^\epsilon dx f( x)\ 2y_m
\int_{-1+\epsilon}^{1-\epsilon} dk_x'\  2\ln\frac{1+x-k'_x}{1-x-k'_x}\\
&=4\int_0^\epsilon dx f( x)\ y_m\int_{\epsilon}^{2-\epsilon} du\  \ln\frac{u+x}{u-x}\\
&\leq 4\int_0^\epsilon dx f( x)\ y_m\int_{\epsilon}^{2-\epsilon} du\ \frac{2x}{u-x}\\
&=8\epsilon^2\sqrt{2\epsilon}\left[\ln\epsilon^{-1}+O(1)\right]\int_0^1 dx f(\epsilon x)x\sqrt x
\end{align}
\section{Appendix B}
We have to estimate
\begin{align}
\label{dvdef}
I(f,g)=\int_{\begin{subarray}{1} \|k\|,\|k'\|\leq 1\\1-k_x,1-k'_x<\epsilon\end{subarray}}dkdk'\ \left(v_{k-k'}f-v_{\tilde k+k'}g\right)
\end{align}
where $\tilde k=(k_x-2,k_y)$ and $f$ and $g$ are positive functions of $1-k_x$ and $1-k_{x'}$.\\
Setting $k_x=1-x$, $k'_x=1-x'$, $k_y=y$,  $k'_y=y'$, and $r_\pm=\sqrt{(x\pm x')^2+(y-y')^2}$, Eq.\ref{dvdef} can be rewritten:
\begin{align}
\nonumber
\nonumber
I(f,g)&=\int_0^\epsilon dx\int_0^\epsilon dx' \int dydy'\left(\frac 1{r_-}f-\frac 1{r_+}g\right)\\
\label{dv1}
&=2\int_0^\epsilon dx\int_x^\epsilon dx' \int dydy'\left(\frac 1{r_-}f-\frac 1{r_+}g\right)
\end{align}
where $y$ and $y'$ must satisfy $(1-x)^2+y^2\leq 1$ and $(1-x')^2+{y'}^{2}\leq 1$.\\
Since $\asinh x\leq \ln 2(x+1)$, the first term in \ref{dv1} is bounded by:
\begin{align}
2\int_0^\epsilon dx\int_x^\epsilon dx' \int dydy'\ \frac 1{r_-}f&=2\int_0^\epsilon dx\int_x^\epsilon dx' f\int_{-y_m}^{y_m} dy\ (\asinh \frac{y'_m+y}{x'- x}+\asinh \frac{y'_m-y}{x'- x})\\
&\leq2\int_0^\epsilon dx\int_x^\epsilon dx' f\int_{-y_m}^{y_m} dy\ 2\asinh \frac{2y'_m}{x'- x}\\
&\leq 4\int_0^\epsilon dx\int_x^\epsilon dx' f2y_m\ln (2+\frac{4y'_m}{x'- x})\\
&\leq 4\epsilon^2\sqrt{2\epsilon}\left[\ln(\epsilon^{-1})+O(1)\right] \int_0^1 dx\int_x^1 dx\,f( \epsilon x, \epsilon x')\sqrt x
\end{align}
On the other hand, using $\asinh x\geq \ln 2x$, the last term of (\ref{dv1}) is:
\begin{align}
2\int_0^\epsilon dx\int_x^\epsilon dx' g\int dydy'\frac 1{r_+}
&=2\int_0^\epsilon dx\int_x^\epsilon dx' g\int_{-y_m}^{y_m} dy\asinh \frac{y'_m-y}{x+x'}+\asinh \frac{y'_m+y}{x+x'}\\
&\geq 2\int_0^\epsilon dx\int_x^\epsilon dx' g\int_{-y_m}^{y_m} dy\ln 4\frac{{y'_m}^2-y^2}{(x+x')^2}\\
&\geq 4\int_0^\epsilon dx\int_x^\epsilon dx' g y_m\left[\ln\epsilon^{-1}+O(1)\right]\\
&\geq 4\epsilon^2\sqrt{2\epsilon}\left[\ln\epsilon^{-1}+O(1)\right]\int_0^1 dx\int_x^1 dx' g(\epsilon x, \epsilon x')\sqrt x
\end{align}
And we have:
\begin{align}
I(f,g)&\leq 4\epsilon^2\sqrt{2\epsilon}\left[\ln\epsilon^{-1}+O(1)\right]\int_0^1 dx\sqrt x\int_x^1 dx' \left(f( \epsilon x, \epsilon x')-g( \epsilon x, \epsilon x')\right)
\end{align}

\section{Appendix C}
Here we provide exact bounds on $I_1$ and $I_2$ given by (\ref{i1i2-1}, \ref{i1i2-2}).\\
In order to estimate $I_2$ we introduce the linear operator $A$ :
\begin{align}
Af(x)&=\frac 1 {2x}\int_x^1f(y)dy+\frac 1 {2x\sqrt x }\int_0^xf(y)\sqrt {y} dy
\end{align}
defined on the Hilbert space of the functions on $[0,1]$ with the scalar product:
\begin{align}
\left<f|g\right>&=\int_0^1 x\sqrt x\,\overline{f(x)}g(x) dx
\end{align}
Then $A$ is a bounded symmetric operator and:
\begin{align}
\label{i2i1}
I_2/I_1=\frac{1}{5\pi}\left(\frac{\left<ab|Aab\right>}{\| b\|^2}-1\right)
\end{align}
The unitary operator $f(x)\rightarrow g(y)=f(e^{-y})e^{-5/4y}$ from $L^2([0,1],x\sqrt x dx)$ onto $L^2([0,+\infty],dx)$ maps
the operator $A$ onto the operator $\tilde A$:
\begin{align}
\tilde Ag(x)&=\frac {e^{-x/4}} 2 \int _0^xe^{y/4}g(y)dy +\frac{e^{x/4}} 2 \int _x^{+\infty}e^{-y/4}g(y)dy
\end{align}
Then 
\begin{align}
\tilde Ae^{ikx}&=\frac{1}{4(1/16+k^2)}e^{ikx}-\frac 1 {1/2+i2k}e^{-x/4}
\end{align}
Thus setting:
\begin{align}
g_k(x)=\frac 1 {|1+i4k|}\left[(1+i4k)e^{ikx}-(1-i4k)e^{-ikx}\right]
\end{align}
$\{g_k\}_{k>0}$ is  a full set of pseudo-eigenvectors satisfying: $$\tilde Ag_k=\frac{1}{4(1/16+k^2)}g_k$$
Thus the spectrum of $\tilde A$ is $(0,4)$ and the spectral measure is purely absolutely continuous;  the largest spectral value is $4$ with a pseudo-eigenvector $g_4(x)=x+4$ corresponding to $f_4(x)=x^{-5/4}(4-\ln x)$.\\
But $\|f_4\|$ is infinite and $f_4$ diverges at $0$. The next step is to choose a family of functions $b_\eta$ such that $a_\eta=\sqrt{1-b_\eta}$ is
defined and $\left<a_\eta b_\eta|Aa_\eta b_\eta\right>/\| b_\eta\|^2$ is close to 4.\\
Thus setting $f_{\eta}(x)=\min(f_4(x),f_4(\eta))$ for $0<\eta\ll 1$, we have:
\begin{align}
\|f_{\eta}\|^2&=-\frac 1 3\left[ \ln^3 \eta-\frac{66} 5 \ln^2 \eta +O(\ln \eta)\right]\\
\left<f_{\eta}|Af_{\eta}\right>&=-\frac 4 3\left[\ln^3 \eta-\frac {41} {5} \ln^2 \eta +O(\ln \eta)\right]
\end{align}
Then:
\begin{align}
\label{Afeta}
\frac{\left<f_{\eta}|Af_{\eta}\right>}{\|f_{\eta}\|^2}&=4-\frac {20}{| \ln \eta|}+O(\ln^{-2} \eta)
\end{align}
Thus $f_\eta$ is a good candidate for the linear part of the problem. Now, by the simple scaling:
\begin{align}
\label{beta}
b_{\eta}(x)=\frac{f_{\eta}(x)}{\sqrt 2f_{\eta}(\eta)}
\end{align}
we get the nonlinear candidate satisfying $b_{\eta}(x)\leq1/\sqrt 2$, $a_{\eta}=\sqrt{1-b_{\eta}^2}$ is well defined, $a_{\eta}(x)\geq1/\sqrt 2$ and $b_{\eta}$ satisfies (\ref{Afeta}).\\
We must now estimate the simultaneous convergence of $I2/I1$ (\ref{i2i1}) and $I_2$ as $\eta$ decreases.\\
\begin{align}
\nonumber
\left<b_{\eta}|Ab_{\eta}\right> -\left<b_{\eta}a_{\eta}|Ab_{\eta}a_{\eta}\right>&=-\left<b_{\eta}-b_{\eta}a_{\eta}|A|b_{\eta}-b_{\eta}a_{\eta}\right>
+2\left<b_{\eta}-b_{\eta}a_{\eta}|A|b_{\eta}\right>\\
\nonumber
&\leq 2\left<b_{\eta}-b_{\eta}a_{\eta}|A|b_{\eta}\right>\\
\nonumber
&= 8\left<b_{\eta}-b_{\eta}a_{\eta}|b_{\eta}\right>+2\left<b_{\eta}-b_{\eta}a_{\eta}|(A-4)b_{\eta}\right>\\
\nonumber
&\leq 8\left<b_{\eta}-b_{\eta}a_{\eta}|b_{\eta}\right>+2\|b_{\eta}-b_{\eta}a_{\eta}\|\| (A-4)b_{\eta}\|\\
\nonumber
&\leq 8\left<b_{\eta}-b_{\eta}a_{\eta}|b_{\eta}\right>+8\|b_{\eta}-b_{\eta}a_{\eta}\|\sqrt{\left< b_{\eta}|(A-4)b_{\eta}\right>}
\end{align}
where:
\begin{align*}
\left<b_{\eta}-b_{\eta}a_{\eta}|b_{\eta}\right>&=\int_0^1b_{\eta}(x)^2\left[1-a_{\eta}(x)\right]x\sqrt xdx\\
&\leq \int_0^1b_{\eta}(x)^2\left[1-a_{\eta}(x)\right]^2x\sqrt{x}dx \sup \frac 1 {1- a_{\eta}}\\
&\leq \|b_{\eta}-b_{\eta}a_{\eta}\|^2 \frac {\sqrt 2}{\sqrt 2-1}\\
\end{align*}
and since $b_{\eta}-b_{\eta}a_{\eta}>0$
\begin{align*}
\|b_{\eta}-b_{\eta}a_{\eta}\|^2&=\|b_{\eta}\|^2-\|b_{\eta}a_{\eta}\|^2-2\left<b_{\eta}-b_{\eta}a_{\eta}|b_{\eta}a_{\eta}\right>\\
&\leq \|b_{\eta}\|^2-\|b_{\eta}a_{\eta}\|^2\\
&= \|b_{\eta}^2\|^2
\end{align*}
By direct computation: $$\|b_{\eta}^2\|^2\leq \|b_{\eta}\|^2 \frac{6}{5|\ln \eta|}$$
for $\eta$ small enough, and thus:
\begin{align}
\label{dbAb}
\frac{\left<b_{\eta}|Ab_{\eta}\right>}{\|b_{\eta}\|^2} -\frac{\left<b_{\eta}a_{\eta}|Ab_{\eta}a_{\eta}\right>}{\|b_{\eta}\|^2} \leq  \frac{8}{|\ln \eta|}\left(\frac {\sqrt 2}{\sqrt 2-1}\frac 6 5+2\sqrt 6\right)
\end{align}
And finally, from (\ref{Afeta}) for $b_{\eta}$ and (\ref{dbAb}) and $I_1= \|b_{\eta}\|^2$,   (\ref{i2i1}) gives:
\begin{align}
I_2/I_1&\geq\frac{1}{5\pi}\left(3-\frac{C}{|\ln \eta|}\right)+O(\ln^{-2} \eta)\\
I2&\leq \frac 2 {15\pi} \eta^{5/2}|\ln^3 \eta|\left[1+O\left(\frac{1}{|\ln \eta|}\right)\right]
\end{align}
where
\begin{align}
C=20+8\left(\frac {\sqrt 2}{\sqrt 2-1}\frac 6 5+2\sqrt 6\right)\approx 92
\end{align}
Choosing $\eta$ sufficiently small, this proves that $\Delta E$ is strictly negative for any $r_s>0$; furthermore choosing $\eta$ to minimize $\Delta E$ (Eq. \ref{DEF1}),  i.e. $9r_s|\ln \eta|^2=2C\pi$, we obtain as $r_s$ goes to 0:
\begin{align}
\Delta E&\lesssim -r_s\exp\left(-\frac {5\pi}{3r_s}+\frac {O(1)} {\sqrt r_s}\right)
\end{align}

\end{document}